# Chirality at two-dimensional surfaces: A perspective from small molecule alcohol assembly on Au(111)

Melissa L. Liriano,[†] Amanda M. Larson,[†] Chiara Gattinoni,[‖] Javier Carrasco,[‡] Ashleigh E. Baber,[†] Emily A. Lewis,[†] Colin J. Murphy,[†] Timothy J. Lawton,[†] Matthew D. Marcinkowski,[†] Andrew J. Therrien,[†] Angelos Michaelides,[‖] E. Charles H. Sykes[*,†]

[†] Department of Chemistry, Tufts University, Medford, Massachusetts 02155, USA
[‡] CIC Energigune, Albert Einstein 48, 01510 Miñano (Álava), Spain
[‖] Thomas Young Centre, London Centre for Nanotechnology and Department of Physics and Astronomy, University College London, London WC1E 6BT, United Kingdom
[*] Corresponding author: charles.sykes@tufts.edu

**ABSTRACT**

The delicate balance between hydrogen bonding and van der Waals interactions determine the stability, structure and chirality of many molecular and supramolecular aggregates weakly adsorbed on solid surfaces. Yet the inherent complexity of these systems makes their experimental study at the molecular level very challenging. In this quest, small alcohols adsorbed on metal surfaces have become a useful model system to gain fundamental insight into the interplay of such molecule-surface and molecule-molecule interactions. Here, through a combination of scanning tunneling microscopy and density functional theory, we compare and contrast the adsorption and self-assembly of a range of small alcohols from methanol to butanol on Au(111). We find that that longer chained alcohols prefer to form zigzag chains held together by extended hydrogen bonded networks between adjacent molecules. When alcohols bind to a metal surface datively via one of the two lone electron pairs of the oxygen atom they become chiral. Therefore, the chain structures are formed by a hydrogen-bonded network between adjacent molecules with alternating adsorbed chirality. These chain structures accommodate longer alkyl tails through larger unit cells, while the position of the hydroxyl group within the alcohol molecule can produce denser unit cells that maximize intermolecular interactions. Interestingly, when intrinsic chirality is introduced into the molecule as in the case of 2-butanol the assembly changes completely and square packing structures with chiral pockets are observed. This is rationalized by the fact that the intrinsic chirality of the molecule directs the chirality of the adsorbed hydroxyl group meaning that heterochiral chain structures cannot form. Overall this study provides a general framework for understanding the effect of simple alcohol molecular adstructures on hydrogen bonded aggregates and paves the way for rationalizing 2D chiral supramolecular assembly.

**I. INTRODUCTION**

The interplay of hydrogen bonding and van der Waals (vdW) interactions is important to DNA base pairing, protein structure and function, as well as an almost endless list of materials properties.[1,2] This delicate interplay underpins many important aspects of molecular recognition and self-assembly, but due to their dynamical nature, the long-range structures arising from these interactions are difficult to study experimentally. Simple model systems are thus necessary in order to study and understand individual properties of these complex structures.[3–5] Surfaces provide a good way to tease out these interactions in detail, especially when looking at the adsorption of molecules on noble metals. It has been shown that a rich variety of structures can form, motivating structural studies for understanding interactions and molecular adsorption in regards to catalysis. For the case of hydrogen bonding, a well-studied system is water, both in bulk and adsorbed on surfaces.[6–11] In particular, low-temperature scanning tunneling microscopy (LT-STM) has proven invaluable in understanding the interaction of water with a range of well-defined surfaces and revealed a myriad of complex molecular and partially dissociated networks.[12–16] Moving beyond water towards more complex structures, alcohols represent one of the simplest systems with which to understand local and long-range structures arising from hydrogen bonding as well as ubiquitous vdW interactions. Indeed, the variety of alcohol structures, with different sized secondary alkyl chains offer the ability to interrogate the effect of molecular size and shape on the properties of self-assembled surface layers. Thus, in this case, hydrogen bonding and vdW forces are probed for different molecular sizes, providing new insight on these important interactions in nature and on their interplay in stabilizing molecular systems.[17–22] Often distinct from bulk material properties, chirality can be important and even the basic 'rules' of hydrogen bonded systems at surfaces can be different, necessitating continued research.[23]

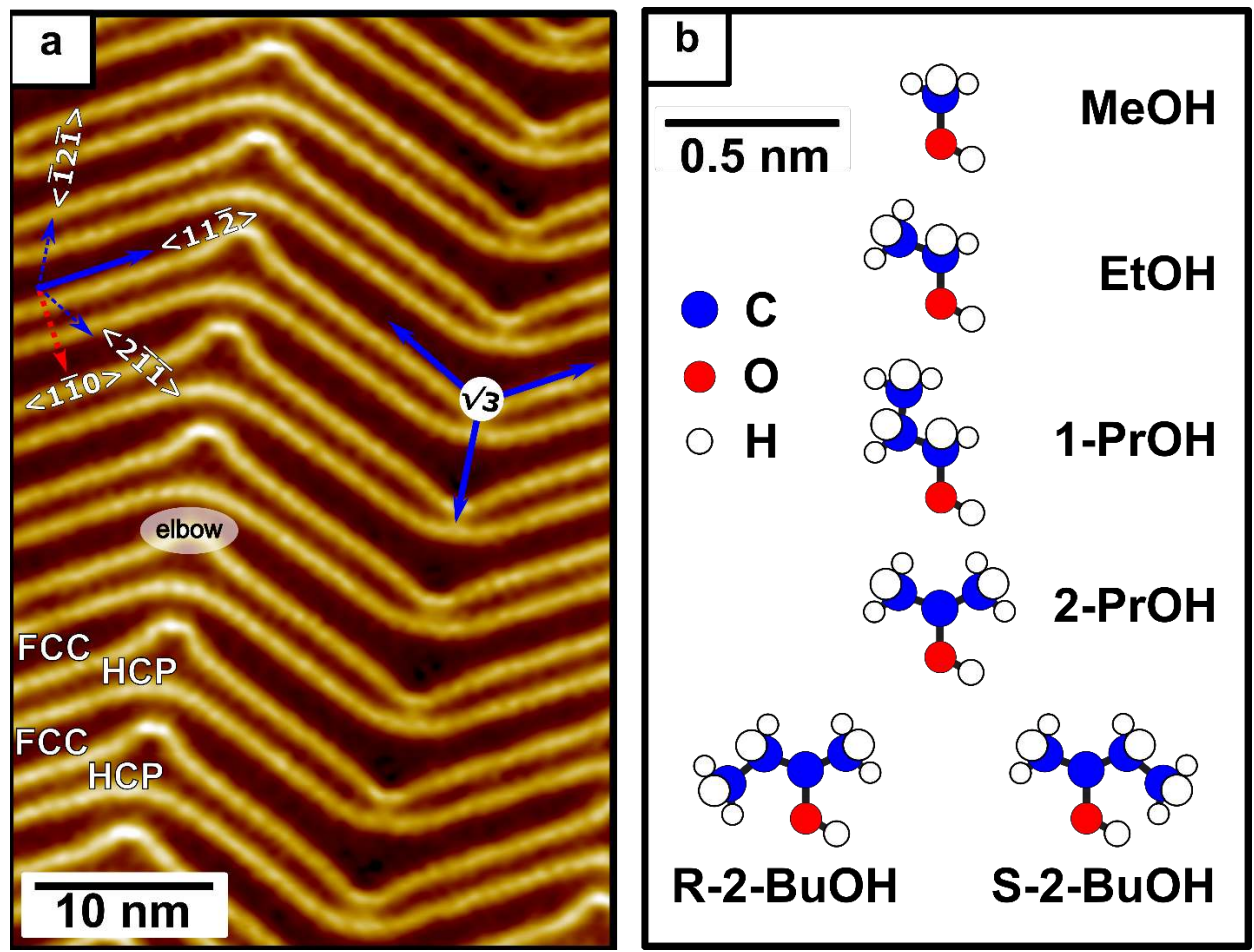

**Figure 1.** (a) LT-STM image of Au(111) herringbone reconstruction. Imaging conditions: 300 mV, 300 pA. (b) Ball-and-stick models for the small aliphatic alcohols studied, with hydrocarbon tail lengths ranging from $C_1$ to $C_4$.

## II. EXPERIMENTAL

**A. STM Experiments**.

All STM experiments were performed on either an Omicron NanoTechnology low temperature (LT) STM or variable-temperature (VT) STM. The base pressure in the LT-STM chamber was 1 x $10^{-11}$ mbar. The Au(111) single crystal sample was cleaned by cycles of $Ar^+$ sputtering (14 µA, 1 kV) and annealing (1000 K). The sample was then transferred into the pre-cooled STM stage with a base temperature of 5 K. The base pressure in the VT-STM chamber at 29 K was < 5 x $10^-$ [11] mbar. The Au(111) single crystal sample was cleaned by cycles of Ar$^+$ sputtering (12 µA, 1 kV) and annealing (1000 K).

Pure MeOH (99.9+%, ultrapure HPLC grade) was purchased from Alfa Aesar. Pure EtOH (99.8%), 1-PrOH (99.9%), 2-PrOH (99.5%), and enantiomers of (*R*)- and (*S*)-2-BuOH (99.9%) were purchased from Sigma Aldrich. Several cycles of freeze-pump-thawing were performed to further purify each alcohol. Submonolayer and monolayer (ML) coverages of each alcohol were deposited through a high-precision leak valve onto the sample which was held at 29 K for all MeOH experiments (VT-STM) or at 5 K (LT-STM) for the EtOH, 1- and 2-PrOH, and (*R*)- and (*S*)-2-BuOH experiments. For MeOH, following a high dose resulting in multilayer surface coverage, a series of anneals ranging between 139-160 K were performed. Annealing to 139 K desorbs all but monolayer coverage of MeOH on the surface, and after annealing to 157 K hexamers are the dominate structure.[17] After each anneal the sample was cooled back down to 29 K and imaged. After dosing EtOH, 1- and 2-PrOH, and (*R*)- and (*S*)-2-BuOH, each system was annealed to 100 K or 120 K, then cooled and imaged at 5 K. Thermal anneals were performed to equilibrate the molecular assemblies. STM images were obtained with Omicron-etched W tips at bias voltages between -1 V and 300 mV and tunneling currents between 5 pA and 500 pA.

**B. Theoretical Methods.**

Density functional theory (DFT) calculations were performed using the optB88-vdW functional[24], a modified version of the non-local vdW-density functional of Dion et al.[25] which has shown good agreement with experimental results in a variety of systems of molecules adsorbed on metals.[19,26–28] The optB88-vdW calculations were carried out self-consistently in VASP using the implementation of Klimeš et al.[29]. Core electrons were replaced by projector augmented wave (PAW) potentials,[30] whereas the valence states were expanded in plane-waves with a cut-off energy of 500 eV. Adsorption calculations of monomers of ethanol were carried out considering (6×6) metal slabs cut along the Au(111) direction consisting of 3 atomic layers thickness and separated by 18 Å of vacuum. In the case of adsorbed chains, a (10×2) unit cell was used. The metal atoms in the bottom layer were fixed to the bulk optB88-vdW optimal positions ($a_{Au}$=4.158 Å) whereas all other atoms (in the substrate and adsorbates) were allowed to relax. We used a Monkhorst-Pack k-points grid[31] of 2×2×1 and 1×6×1 for the (6×6) and (10×2) unit cells, respectively. A dipole correction along the direction perpendicular to the metal surface was applied,[32,33] and geometry optimizations were performed with a residual force threshold of 0.025 eV/Å. STM images were simulated using the Tersoff-Hamann approach,[34] with a voltage of −100 mV and at a height of ~6.5 Å above the metal surface. The relative stability of different trial structures was assessed *via* their adsorption energy, $E_{ads}$, defined as: $E_{ads}$ = ($E_{system}$ - $E_{Au(111)}$ - n× $E_{mol}$ )/n, where $E_{system}$, $E_{Au(111)}$ and $E_{mol}$ are the total energy of, respectively, the whole adsorbed system, the Au(111) slab and an alcohol molecule in the gas phase. As discussed in the next section, Au(111) undergoes a well-known 'herringbone' reconstruction. We did not attempt to model this in our simulations and used a relaxed but otherwise unreconstructed Au(111) surface.

## III. RESULTS AND DISCUSSION

### A. Au(111) Reconstruction

In order to discuss our data, a brief description of the Au(111) reconstruction is necessary. The reconstruction of a clean Au(111) surface occurs naturally in order to relieve lateral strain due to crowding on the first atomic layer. More specifically, the unit cell of the reconstructed surface consists of 23 atoms sitting on 22 bulk lattice sites, creating a long-range elastic lattice strain in the atomic surface layer. The surface adopts a 22 × √3 arrangement, often referred to as the herringbone (HB) reconstruction, which involves a 4.5% contraction along a close-packed, or $[1\bar{1}0]$, direction forming stacking faults consisting of wider FCC and narrower HCP packed regions.[35] Meanwhile, the low symmetry $[1\bar{1}2]$, or √3 direction, that lies perpendicular to the compressed $[1\bar{1}0]$ axis is completely uncompressed. The two other √3 directions are oriented 30° from the compressed close-packed direction and are partially compressed by 3.9%.[36] The FCC-HCP stacking transitions are separated by soliton walls, which appear as pairwise corrugation lines. On a clean Au(111) surface, the distance between neighboring pairs of solitons is 6.3 nm. Au atoms on the first atomic layer rest in a variety of sites, with FCC and HCP atoms sitting on three-fold hollow sites while soliton wall atoms sit topographically higher on 'quasibridge' sites.[35] STM images of the clean Au(111) HB reconstruction show that the soliton walls appear as bright zigzag lines that run in three equivalent √3-directions to relieve strain isotropically, as shown in Figure 1a.

**B. Methanol on Au(111)**

High-resolution VT-STM images revealed that MeOH formed at least two types of hydrogen bonded structures (hexamers and zigzag chains), and that the proportion of each hydrogen bonded structure was dependent on MeOH surface coverage.[17] The first hydrogen bonded structure consists of cyclic hexamers found on FCC and HCP sites of the Au reconstruction, for 0.07 ML coverage, Figure 2a-b. These MeOH hexamers did not coalesce, indicating repulsive interactions between the hydrogen bonded cyclic networks.[17] Hexamer formation is driven by hydrogen bonding interactions between adjacent molecules, similar to our second structure, the zigzag chain. At 0.4 - 0.5 ML surface concentrations, hydrogen bonded structures consisting of winding chains and extended rings evolve based on a zigzag chain motif. The zigzag chain features preferred the compressed √3 surface symmetry directions, with experimental measurements in agreement with the Au-Au atom spacing in this direction for MeOH molecules within the chain, indicating that growth was epitaxial on the underlying surface.[17] In addition, the gap between chain pairs increased from $0.77 \pm 0.03$ nm to $1.5 \pm 0.1$ nm, after a 148 K anneal (0.5 ML). A slightly higher anneal to 153 K (0.4 ML), as shown in Figures 2c-e, resulted in larger chain-chain separations of $1.8 \pm 0.1$ nm. This increase in the gap or space between chain pairs that is observed with subsequent anneals (as coverage decreases) indicates that MeOH chains repel one another. At 1 ML, the dominant features are zigzag, pairwise chains, as shown in Figure 2f. This third structure utilizes the zigzag chain motif to pair chains into a long-range ordered monolayer. The 11× √3 unit cell for this extended chain structure contains MeOH chains running parallel to the two partially compressed √3 symmetry directions of the underlying surface. The distance between equivalent points on pairwise chains is the unit cell length in the close-packed direction, but the distance between adjacent pairwise chains is $1.55 \pm 0.05$ nm[17], see green line in Figure 2g. Experimental measurements confirmed that molecule-molecule spacing was $0.48 \pm 0.2$ nm in good agreement with the Au-Au atom spacing in the compressed √3 directions, $[\bar{1}2\bar{1}]$ or $[2\bar{1}\bar{1}]$. The underlying HB reconstruction remained unperturbed, an indication that the MeOH-Au interaction is relatively weak.[17,37]

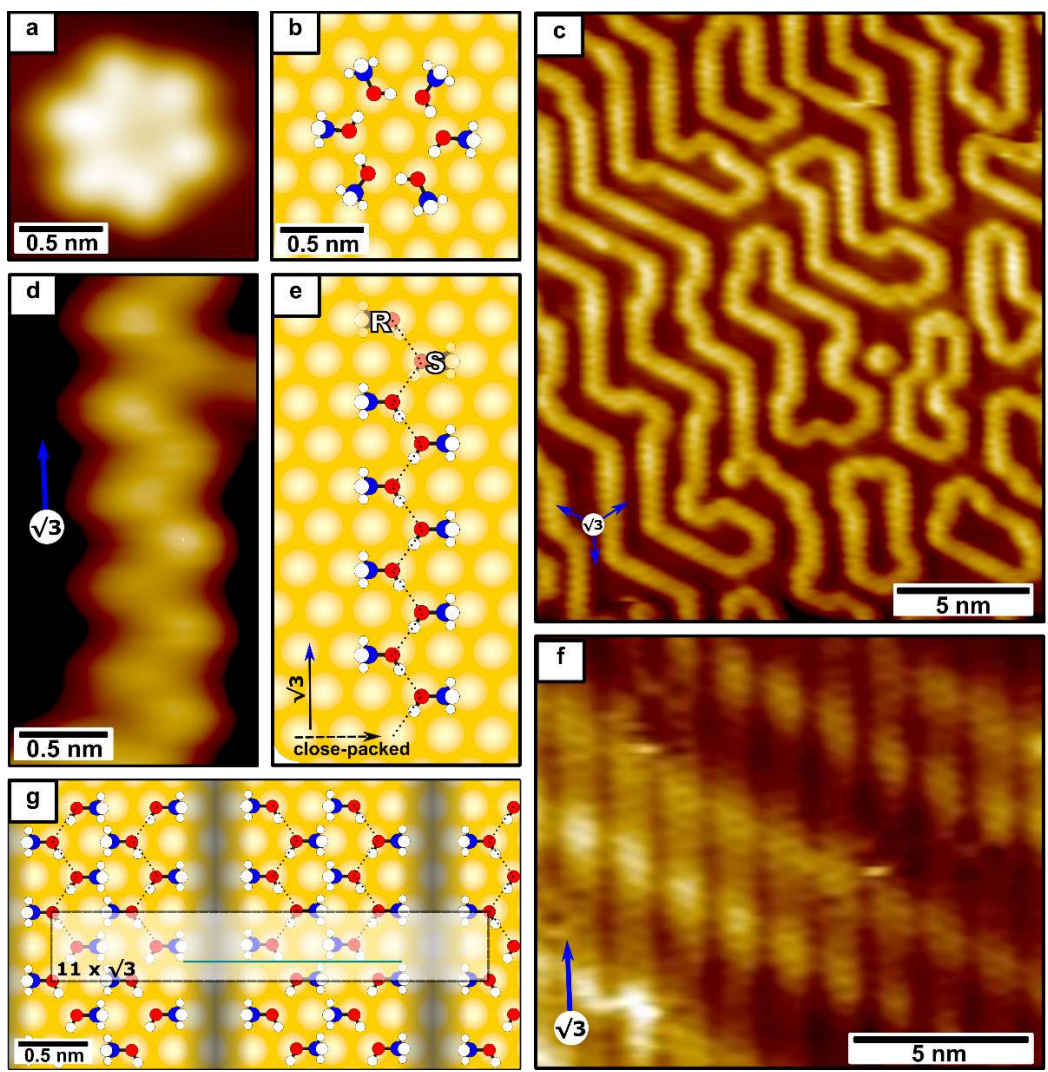

**Figure 2**. (a) STM image of single MeOH hexamer taken with LT-STM at 5 K, scanning conditions: -100 mV, 30 pA. (b) Schematic of MeOH homochiral hexamer. (c) VT-STM image of 0.4 ML of MeOH zigzag chains on Au(111) after a 153 K anneal. (d) Zoomed-in VT-STM image of a single MeOH zigzag chain. Images acquired at 29 K/ Scanning conditions: -1 V, 5 pA. (e) Schematic showing proposed structure for MeOH hydrogen-bonded zigzag chains. Molecules of one point chirality are on the right, the other chirality is on the left of the dotted line bisecting the zigzag chain. (f) STM image of approximately ML coverage MeOH. (g) Schematic showing pairwise chain structure of ML coverage (11 × √3 unit cell); Green line = distance between adjacent pairwise chains.

The proposed models of MeOH assembly on Au(111) have three important structural implications. The first is that the preferred binding site of each methanol molecule is atop or near a Au atom and binding occurs through a lone electron pair of the oxygen atom. The second is that molecular adsorption gives rise to a type of surface-bound chirality, termed 'point chirality', where a chiral center develops at the oxygen atom, with the Au surface as the highest priority group. Each MeOH surface-bound enantiomer is distinguished by the direction that the OH group points, as illustrated in Figure 2e. The third is that the chirality of each surface-coverage-dependent MeOH structure differs. For instance, in Figure 2e, the zigzag chains are composed of MeOH molecules hydrogen bonded to adjacent molecules with alternating surface-bound chirality, which was confirmed by DFT calculations and STM simulations.[18,19] MeOH zigzag chains, are therefore, heterochiral structures. MeOH hexamers are composed of six hydrogen bonded molecules all with the same surface-bound chirality, which enables them to arrange in a cyclic manner, Figure 2b. Two types of homochiral hexamers were observed, consisting of six surface-bound MeOH molecules, one with a clockwise rotation and a second with a counter-clockwise rotation.[18] Experimental measurements found that the hexamers were rotated ± 5° from the high symmetry close-packed direction of the underlying Au surface, rendering the structure themselves chiral. Since each surface-bound enantiomer is equally stable on the surface, both chiral hexamers exist in equal concentrations and therefore the overall system is achiral. DFT calculations confirmed that the direction of the OH group of each MeOH enantiomer dictated the direction the hexamer was rotated relative to the close-packed surface directions.[18] Therefore, cluster rotation is dictated by the direction of the hydrogen bonded network, directly associated with the point chirality of the MeOH molecules themselves.

**C. Ethanol on Au(111)**

Our high-resolution images show EtOH molecules on the Au(111) surface at varying concentrations. At low surface coverages, STM images feature single (1D) chains adsorbed predominantly on the FCC regions of the HB reconstruction, as shown in Figure 3. Like MeOH at higher coverages, the single EtOH chains run parallel to the √3, or the $\langle 11\bar{2} \rangle$, directions of the underlying substrate. A minority species consisting of five lobed-cyclic structures are found exclusively at the HB reconstruction elbows at which surface defects comprised of edge dislocations at the vertices of the V-shaped soliton walls exist, see inset of Figure 3a. Due to under-coordinated Au atoms, the edge dislocations can bind molecules more strongly and nucleate structures that do not necessarily represent the most stable structures observed or theoretically predicted on "flat" surface regions. For this reason, pentameric structures observed on HB elbows will not be considered in this discussion.

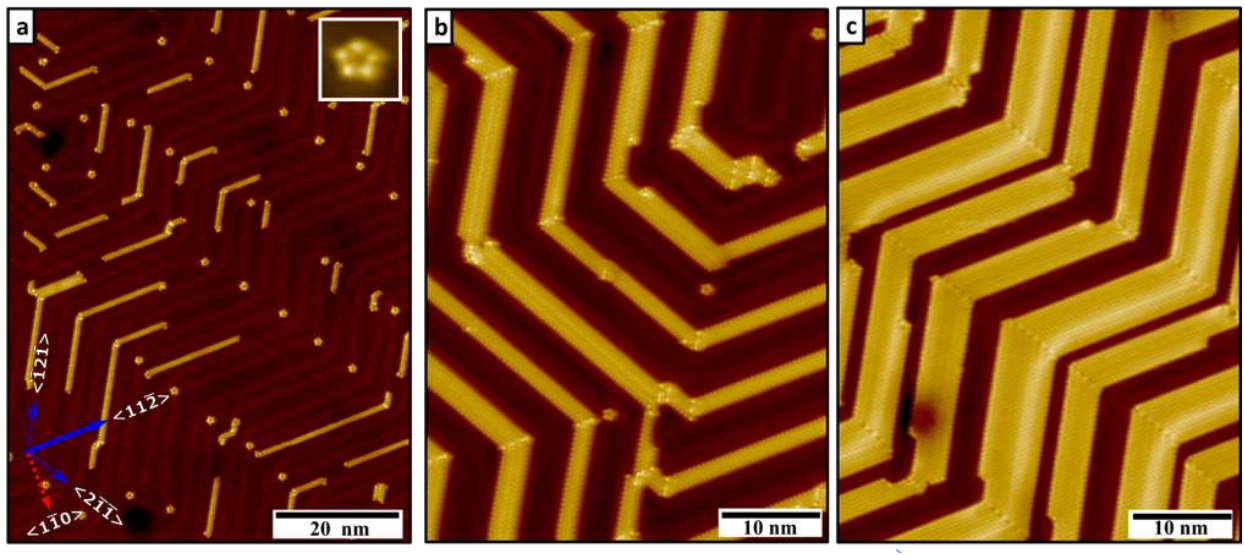

**Figure 3**. STM images of various concentrations of EtOH on Au(111). Images acquired at 5 K after ≥ 80 K anneal. (a) Low coverage of EtOH after an ~80 K anneal showing that the dominant features are zigzag chains in FCC regions and pentamers on elbow edge dislocations (2.5 nm wide inset). (b) Submonolayer coverage of EtOH, image acquired after an ~ 80 K anneal showing that the dominant structures are double chains, with pentamers as the minority species found exclusively at edge dislocations of the HB reconstruction. (c) Near monolayer coverage of EtOH, image acquired after a 120 K anneal showing that the only structures observed are densely packed chains. All structures run in √3, or the $\langle 11\bar{2}\rangle$, symmetry directions of the underlying surface. Scanning conditions: +40-50 mV, 20-500 pA.

One key observation is that EtOH preferentially forms chains at low concentrations, whereas the dominant features of MeOH at comparable coverages were hydrogen-bonded chiral hexamers composed of six methanol molecules.[17,18] Our STM data showed that chains continue to be the dominant feature from very dilute EtOH surface concentrations up to near ML surface coverages, as shown in Figures 3a-c. Conversely, in the MeOH system, it was observed that with increasing surface coverage the thermodynamically stable adsorbed species went from exclusively homochiral hexamers to heterochiral zigzag chain structures at the ML regime.[17,19]

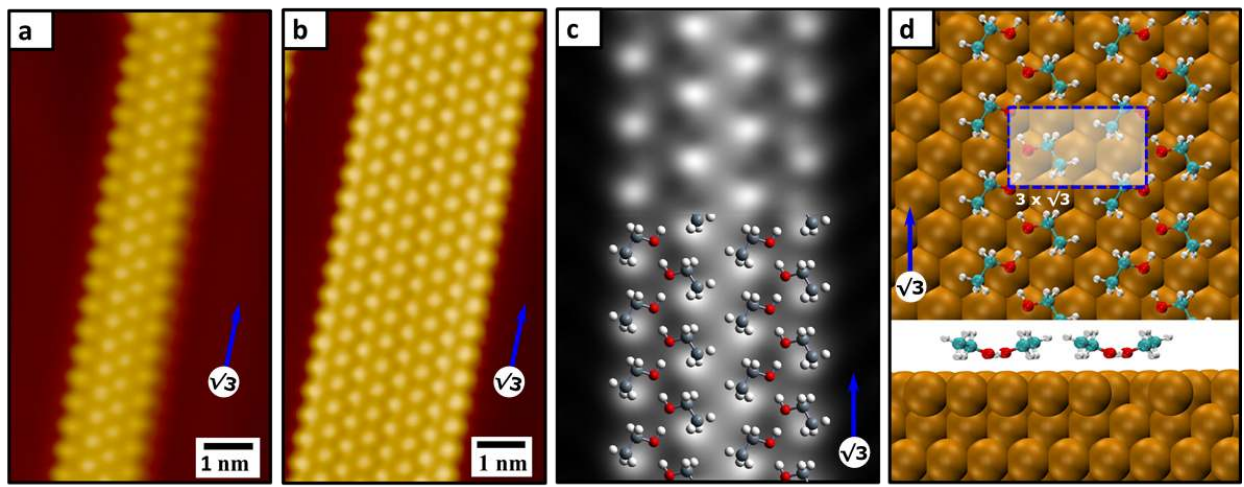

**Figure 4**. (a-b) STM images of EtOH double and quadruple chains observed across varying surface concentrations on Au(111); image was acquired at 5 K after a 120 K anneal. Scanning conditions: 50 mV, 20-25 pA. (c) DFT STM simulated image of an EtOH double chain, which agrees well with experimental observations (at a height of 6.5Å, V= -0.1V); the bottom half of the simulated image has a schematic of the proposed zigzag chain structure overlaid. (d) DFT calculated structure of an EtOH double chain, showing that each EtOH molecule binds to near atop sites on the Au(111) lattice and that hydrogen bonding and vdW interactions drive chain assembly. The blue dotted lines highlight the proposed unit cell, 3x√3. A side view of the molecules is also given in the lower portion of the panel.

With increasing EtOH coverage, the zigzag chain width increases from single to multiple wide chains. Our proposed model for the structure of EtOH chains resembles the zigzag structures observed for MeOH on Au. DFT calculations predict that, similar to MeOH, upon adsorption, each EtOH molecule binds atop a Au atom through an oxygen lone pair, with the OH bond and methyl group lying almost parallel to the surface. Consequently, we observe again the expression of a surface-bound chiral effect or point chirality developed at the adsorbed oxygen center. And the EtOH chain structures are composed of molecules with alternating surface-bound chiralities, hydrogen-bonded to one another, forming an internal zigzag pattern as illustrated in the DFT calculated structure in Figure 4d. Notice that this zigzag pattern is similar to that found for MeOH chains (Figure 2c). The DFT simulated STM image, in Figure 4c, indicates that the bright protrusions observed with STM correspond to the methyl groups. The proposed structure also suggests that lateral growth for double, triple and larger widths are stabilized by vdW interactions between neighboring molecules.

Distance measurements performed on high-resolution STM images of EtOH chains composed of 2 or more rows revealed that the average lobe-lobe distance in the √3 direction was 0.51 ± 0.02 nm. This distance corresponds to equivalent oxygen atoms in adjacent EtOH molecules of the same chirality, which agrees well with the theoretical measurement for the Au-Au spacing of 0.50 nm in the uncompressed √3 direction that runs perpendicular to the closed-packed or $[1\bar{1}0]$ compressed axes. When identifying equivalent adsorption sites for EtOH chains composed of four or more rows we found that the average lateral distance between equivalent locations on adjacent zigzag chains was 0.83 ± 0.04 nm. This experimental measurement agrees well with the proposed molecular arrangement for which the distance between equivalent oxygens is predicted to be 0.83 nm in the close-packed, or $[1\bar{1}0]$, compressed high symmetry direction, as shown in the unit cell (3 × √3) highlighted in Figure 4d.

At low EtOH concentrations, the heterochiral chain structures are single zigzag chains; and as the surface coverage increases, larger chain structures begin to emerge as lateral growth is stabilized by the vdW interactions between adjacent hydrocarbon chains. Interestingly, with increased EtOH surface coverage, the chain spacing remained constant, and upon comparison with the surface area per molecule of the MeOH (11 × √3) pairwise chain superstructure, this indicates that EtOH chains do not experience the same degree of repulsive interactions as observed for MeOH. This finding suggests that the attractive vdW interactions between the longer hydrocarbon tail of EtOH as compared with MeOH counteract the effect of repulsion between the adsorbed heterochiral EtOH species.

**D. 1- and 2-Propanol on Au(111)**

In this section, we present new results from experiment using the LT-STM to study the assembly of slightly larger alcohols, 1- and 2-propanol on Au(111) at 5 K. The high-resolution STM images of 1-PrOH in Figures 5a-b reveal that at low surface concentrations zigzag chains form on Au(111) exclusively. This trend persists as the surface coverage is increased and the 1 ML regime is reached.

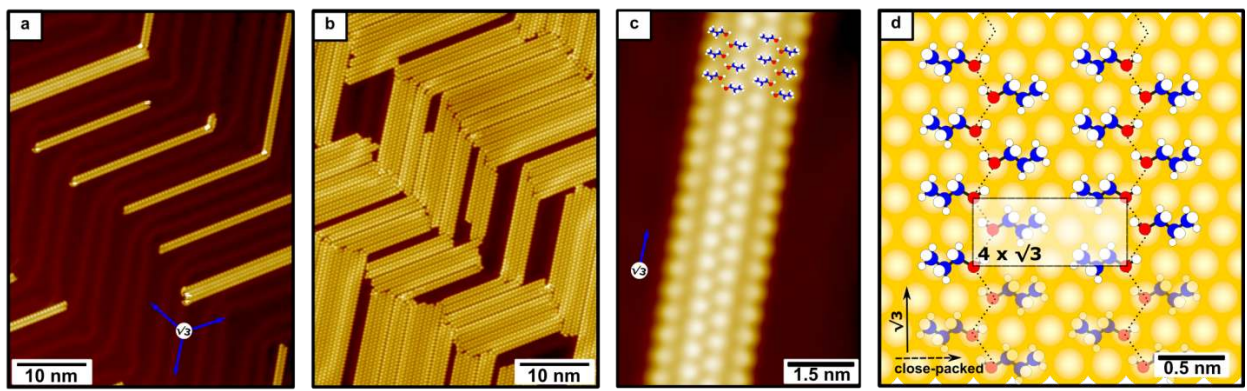

**Figure 5**. STM images of various concentrations of 1-PrOH on Au(111) acquired at 5 K after annealing. Scanning conditions: 10-300 mV, 400 pA (a) Low surface concentration of 1-PrOH. (b) Near monolayer coverage of 1-PrOH. (c) High-resolution STM image of a 1-PrOH double chain, with schematic showing proposed assembly of 1-PrOH molecules overlaid on top portion of image. (d) Schematic of proposed double chain 1-PrOH structure. The opaque box highlights the proposed unit cell, 4 × √3.

Again, we find that the 1-PrOH chains are aligned with the √3 directions of the underlying substrate. At low surface coverages, they are concentrated on the FCC regions of the Au reconstruction and then proceed to cover the whole surface when reaching ML coverages. The average distance between equivalent molecules or lobes in the √3 direction is 0.51 ± 0.01 nm, which is consistent with the Au-Au spacing of 0.50 nm in the uncompressed √3 axes. The distance between equivalent molecules or lobes in the close-packed direction is slightly larger than was measured for EtOH with an average of 1.14 ± 0.03 nm, which is within range for what is predicted for 4 Au-Au distances in the $[1\bar{1}0]$ or close-packed symmetry direction (1.10 nm). The slightly larger lobe-lobe distance may indicate that the 1-PrOH molecules are not sitting exactly atop but instead slightly off or near the Au atop site. This slight increase can be rationalized by the additional carbon present in 1-PrOH, requiring more space for each molecule within the unit cell to minimize steric hindrance and/or repulsions between hydrocarbon tails in adjacent rows. Figure 5d shows a schematic representation of the proposed structure for the 1-PrOH zigzag chain, with the larger unit cell, 4 × √3, highlighted. The proposed model for the structure of 1-PrOH zigzag chains is reminiscent of the chain model determined for 1-PrOH crystals resolved using x-ray diffraction at 248 K[38]. Furthermore, like the proposed model for the zigzag chains on Au, the chain model for 1-PrOH crystal structures indicates that the $(CH_2)_2CH_3$ hydrocarbon tails of neighboring molecules are positioned parallel to one another.[38] The proposed structure for 1-PrOH on Au(111) conjects that, like the smaller alcohols discussed, each 1-PrOH molecule binds to the surface via one lone pair of the oxygen atom, which upon adsorption exhibits point chirality.

To investigate how the hydrogen bonding, vdW interactions, and self-assembly on the surface would be impacted by changing the position of the OH functional group, 2-PrOH was deposited and imaged using the same conditions. High-resolution STM images, shown in Figure 6a-b, confirm that the formation of zigzag chains is still the most energetically preferred structure across all surface coverages explored. As with the $C_1$-$C_3$ alcohols discussed, 2-PrOH chains are found predominantly on the wider FCC sites of the reconstruction at low coverages and cover the whole surface with increasing concentrations with orientations aligned along the √3 directions of the Au surface. A closer look at these chain structures, however, reveals some variation in the assembly pattern. The zoomed-in STM image in Figure 6c shows that each 2-PrOH molecule appears as a pear-shaped protrusion with a dimmer portion facing inward towards the chain. This differs from the more spherical appearance of MeOH, EtOH, and 1-PrOH observed in the STM images in Figures 2-5. The dim-bright pear-shape appearance of 2-PrOH molecules remains independent of bias voltages and current conditions or tip states.

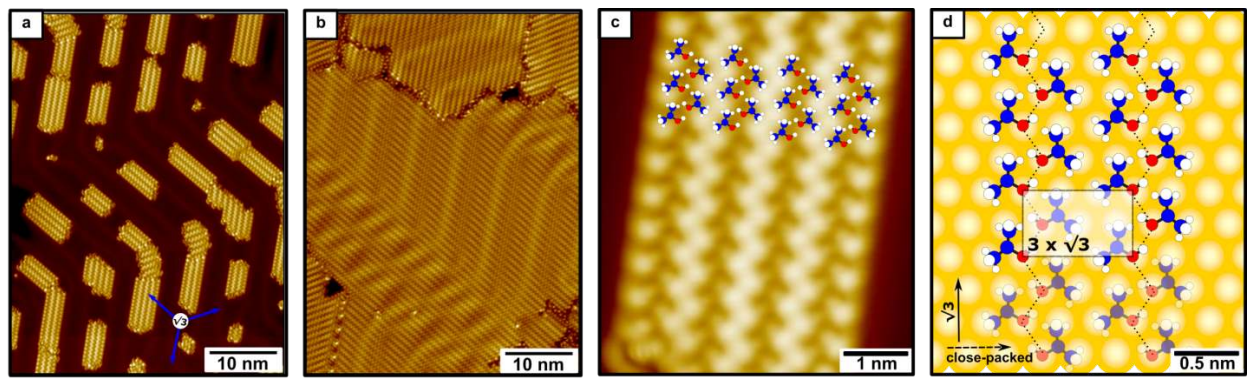

**Figure 6**. STM images of various concentrations of 2-PrOH on Au(111) acquired at 5 K, after annealing. All structures run in √3 symmetry direction of the underlying surface. Scanning conditions: 50 mV, 450-500 pA (a) Low surface concentration of 2-PrOH. (b) Near monolayer coverage of 2-PrOH. (c) High-resolution, zoomed-in STM image of a 2-PrOH quadruple chain, with schematic showing proposed assembly of 2-PrOH molecules overlaid on top portion of image. (d) Schematic of proposed double chain 2-PrOH structure. The opaque box highlights the proposed unit cell, 3 × √3.

Careful measurements of the lobe-lobe distance between equivalent units in the √3 direction results in a 0.52 ± 0.02 nm distance, consistent with the uncompressed Au-Au √3 spacing of 0.50 nm. Further measurements of the distance between equivalent lobes in the close-packed direction reveals that this spacing is only 0.86 ± 0.02 nm, which is smaller than the lateral chain distance of 1-PrOH at 1.14 ± 0.03 nm, but agrees with the 3 Au-Au distances in the closed-packed direction of 0.83 nm. These lobe-lobe distances in 2-PrOH are more comparable with those predicted for the smaller alcohol, EtOH.

The proposed model developed for the 2-PrOH chain structure is shown in Figure 6d. Similar to the proposed models for previous $C_1$-$C_3$ alcohols, 2-PrOH molecules are speculated to have surface-bound chirality upon binding on Au atop sites via an oxygen lone pair and that chain structures are heterochiral species. Positioning the hydroxyl group in the center gives the inner 2-PrOH molecules two points of lateral interaction with neighboring molecules. In addition, since each side of the hydroxyl group is only extended by one carbon atom, each point of interaction with neighboring molecules via vdW interactions between hydrocarbons requires a similar amount of spacing as is required for the smaller EtOH molecules. This molecular arrangement allows for hydrogen bonding interactions to still occur and also maximizes vdW interactions. The proposed unit cell is 3 × √3, which is denser than its structural isomer, 1-PrOH, but similar to that of the smaller EtOH molecule.

**E. (*R*)- and (*S*)-2-BuOH on Au(111)**

Thus far, from our observations for MeOH, EtOH and PrOH we now have a better understanding of how shifting the balance between hydrogen bonding and vdW interactions by increasing hydrocarbon chain length affects self-assembly on an inert metal support. And, in particular, how this balance impacts molecular packing density. Furthermore, all of the $C_1$-$C_3$ alcohols investigated thus far are achiral in the gas phase, with point chirality expressed only upon adsorption due to the development of a stereogenic center around the oxygen atom. So, the presence of equal amounts of surface-bound enantiomers gives rise to homochiral hexamers and heterochiral zigzag chains of these alcohols. It would also be interesting to investigate the impact of intrinsic chirality of the alcohol molecule on surface-bound chirality, molecular packing, and self-assembled overlayers. To this end we compare the adsorption of the simplest aliphatic chiral alcohol, (*R*)- and (*S*)-2-BuOH, on Au(111).[20]

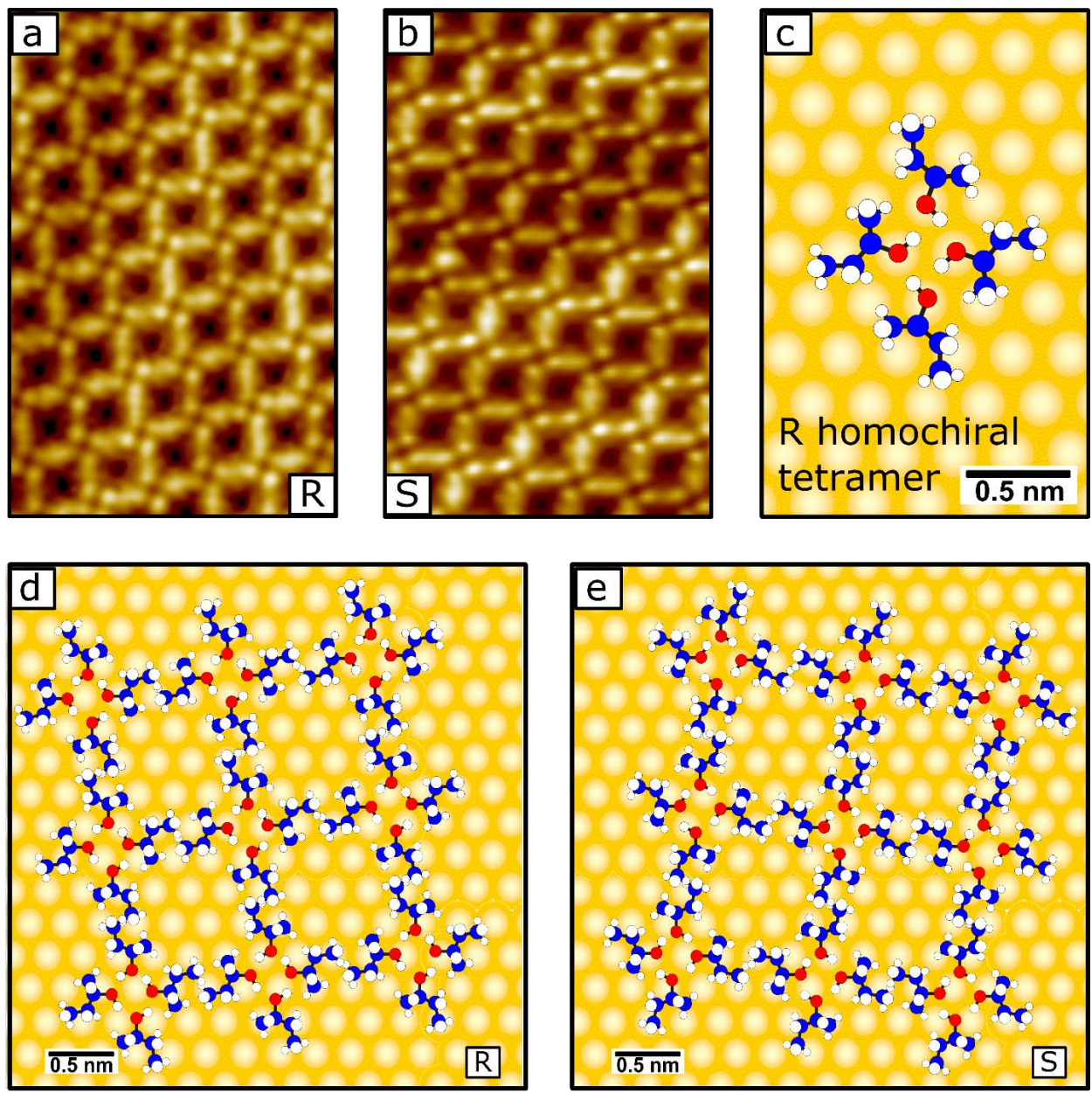

**Figure 7**. STM images of 2-BuOH on Au(111) acquired at 5 K, after a 100 K anneal; images are rotated such that a close-packed direction is horizontal with a vertical mirror plane in the √3 direction. (a) R-2-BuOH chiral square domain. Scanning conditions: 40 mV, 350 pA. (b) S-2-BuOH enantiodomain. Scanning conditions: -100 mV, 50 pA. (c) Schematic of the proposed structure for an R-2-BuOH hydrogen-bonded tetramer. 2-BuOH molecules bind to the Au surface via an oxygen lone pair with the OH bond almost parallel to the surface. (d) Proposed chiral pore structure consisting of nine R-2-BuOH hydrogen-bonded tetramers. (e) Proposed chiral pore structure consisting of nine S-2-BuOH hydrogen-bonded tetramers.

High-resolution STM images revealed the formation of a continuous molecular film over the Au surface consisting of a series of connected square pockets, Figure 7a-b. The central unit of each square pocket was a smaller 4-lobed unit confirmed through DFT calculations to be a hydrogen bonded 2-BuOH tetramer, Figure 7c. Contrary to the 3-fold symmetry of the Au(111) surface, it was found that the formation of tetramers was more energetically preferred than dimers and trimers.[20] DFT calculations also found that hydrogen bonded tetramers were equally as stable as hydrogen bonded hexamers but hexameric units were not observed in the surface coverages explored. The DFT calculated structure for the hydrogen bonded tetramer shows that each 2-BuOH molecule again binds to preferred Au atop sites, with the OH group almost parallel to the surface while the hydrocarbon tail tilts away from the surface.[20]

Similar to the achiral $C_1$-$C_3$ alcohols investigated before, point chirality develops when the chiral 2-BuOH molecule adsorbs on the Au atom via the oxygen lone pair. As such, the 2-BuOH molecules can have both intrinsic and surface-bound chiral centers. In addition, tetramers are composed of 2-BuOH molecules with the same surface-bound chirality, and therefore are homochiral hydrogen bonded species. Each tetramer has a slight rotation, relative to the closed-packed symmetry direction of the Au surface, as was observed with the chiral MeOH hexamers.[17] Specific to the 2-BuOH system, homochiral tetramers orientate in only one rotational offset from closed-packed directions, at - 25˚ and + 25˚ from √3 directions, for (R)-2-BuOH and (S)-2-BuOH, respectively.[20] Each tetramer exists in three different orientations, consistent with the symmetry of the underlying substrate. With the presence of only one type of 2-BuOH tetramer for each chiral species and the homochirality of all tetramers, there exists only one surface-bound enantiomer for each system.[20] The important consequence of this observation is that intrinsic chirality is transferred to the surface-bound chirality inducing enantiospecific adsorption, making only one surface-bound enantiomer energetically preferred. This is supported by DFT calculations which show that in the (R)-2-BuOH system one surface-bound diastereomer molecule is 21 meV more stable, resulting in the increased stability of one type of homochiral hydrogen bonded tetramer by 35 meV.[20] Within a given 2-BuOH system, each 2-BuOH molecule has the same point chirality at the O center, negating the feasibility of forming heterochiral zigzag chains to satisfy the hydrogen bonding requirements; only homochiral cyclic clusters are energetically favorable and thus observed.

The model for the formation of square 2-BuOH films proposed involves hydrogen bonded, homochiral tetramers interacting with neighboring molecules of like-rotation via vdW interactions between hydrocarbon tails to form extended interconnected square pockets.[20] Since the hydrogen bonded, homochiral tetramers exist in three rotations, extended square domains were predicted and experimentally observed to exist in three different orientations, each rotated 120° relative to one another. The three rotational domains of the (*R*)-2-BuOH/Au system were found to be mirror images or enantiodomains of the three rotational domains found in the (*S*)-2-BuOH/Au system, Figure 7d-e.[20] At monolayer coverage, all three rotational domains form a continuous film, with domain boundaries appearing as dark lines. The soliton walls of the herringbone reconstruction are visible and unperturbed by the 2-BuOH monolayer. The 2-BuOH films that cover the Au surface are, therefore, chiral overlayers and the Au(111) surface is said to be chirally modified.

## IV. SUMMARY AND CONCLUSIONS

A combination of VT-STM, LT-STM, and DFT calculations were brought together to better understand the molecular self-assembly of a set of small aliphatic $C_1$-$C_4$ alcohols on the Au(111) surface at cryogenic temperatures. Insights on the adsorbed structures, especially their chirality and into the interplay between hydrogen bonding and vdW interactions were obtained. In all cases, the Au(111) herringbone reconstruction remains unperturbed; indicating that alcohol binding interactions with the Au surface are relatively weak, and structure is dictated through the balance between hydrogen bonding and dispersion forces. The long-range structures arising from these interactions were rigorously studied with STM and we discovered the propensity of alcohols to form long 1-D hydrogen bonded chains when feasible.

Structural models and DFT calculated structures predict that alcohol molecules (methanol, ethanol and 2-butanol) bind to the surface through an oxygen lone pair at or near Au atop sites.[18-20] This adsorption geometry leads to the development of a stereogenic center at the oxygen atom, through which surface induced chirality develops. Statistically, for achiral alcohols (MeOH, EtOH, 1-PrOH and 2-PrOH), both surface-bound enantiomers exist with equal probability, with no net surface chirality. Therefore, the zigzag chain structures are formed by a hydrogen-bonded network between adjacent molecules with alternating adsorbed chiralities. DFT and STM data on 2-BuOH revealed that the intrinsic molecular chirality was preserved and transferred to the surface-bound chirality, inducing enantiospecific adsorption. As a consequence, heterochiral zigzag chain formation (as seen with the smaller alcohols) is unfeasible whereas homochiral clustering into hydrogen bonded tetramers is energetically preferred. In short, when adding intrinsic chirality in the molecular backbone we observe a transfer of chirality resulting in enantiospecific adsorption, which in turn leads to homochiral long-range ordering.

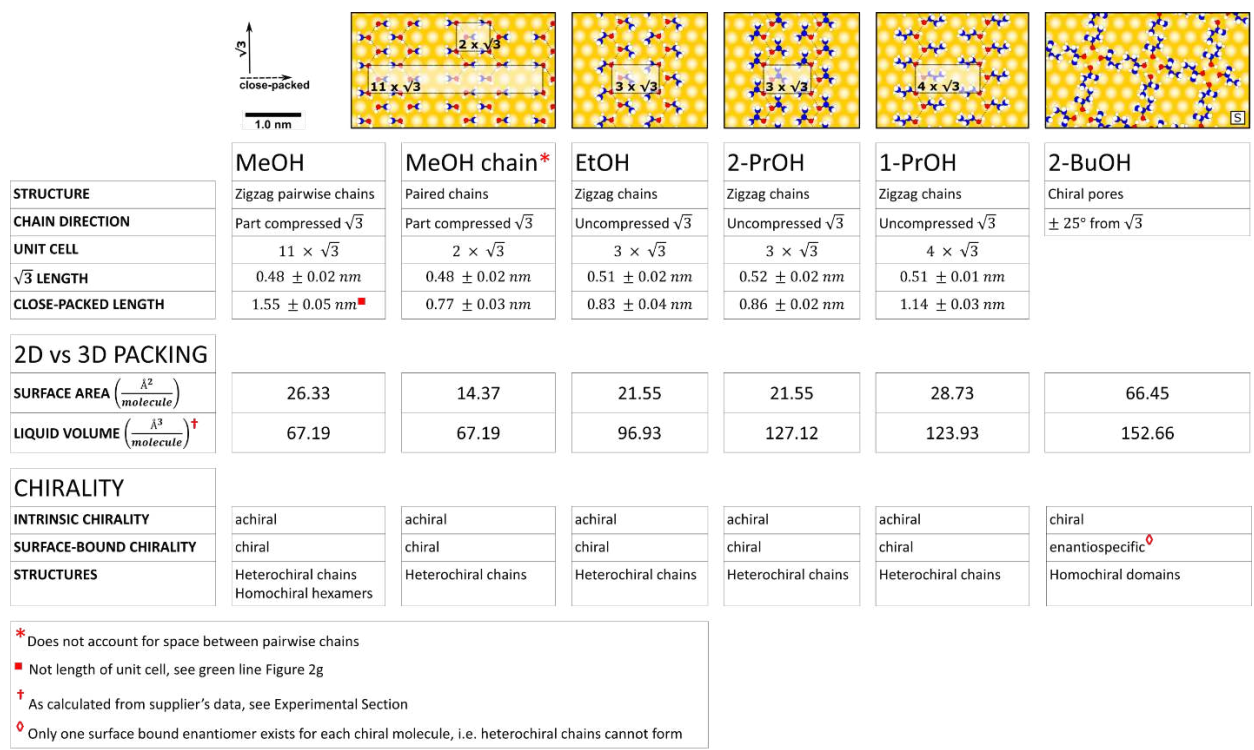

**Figure 8**. Comparison of structures of small alcohols on Au(111); emphasizing differences and similarities between zigzag chain unit cells, packing structures, and chirality.

Experimental data revealed that with increasing chain length, an imbalance in the competing hydrogen bonding-vdW intermolecular interactions occurs impacting cluster formation and growth. More specifically, at dilute surface concentrations the energetically preferred structures of the simplest aliphatic alcohol, MeOH, are two chiralities of homochiral MeOH hexamers held together via hydrogen bonding interactions. Meanwhile, at comparable surface concentrations of EtOH, the most stable structures are zigzag heterochiral chains, with hydrogen bonding dominating interactions between the hydrophilic OH---O domains while lateral vdW interactions between parallel outer hydrophobic carbon tails drives the formation of 1D heterochiral chains. At low surface coverages, lengthening the hydrocarbon tail by just one carbon atom, specifically when going from MeOH to EtOH, is enough to shift the balance between intermolecular interactions towards vdW forces enabling heterochiral chain structures to be more energetically stable than small homochiral hydrogen bonded clusters. Interestingly, with increasing MeOH coverage zigzag chains are formed and at ML coverages resemble those of the other straight chain alcohols. For larger alcohols, 1-PrOH and 2-PrOH, heterochiral chain structures continue to be the most stable structures, exclusively, at all coverages explored. As expected, the molecular packing or unit cell for proposed chain structures for increasingly larger alcohols on Au(111) also grow, with a unit cell of 3x√3 for EtOH chains vs. a larger 4x√3 cell for 1-PrOH chains. Surprisingly, when changing the position of the OH functional group the packing becomes denser for 2-PrOH zigzag chains with a smaller unit cell of 3x√3 than 1-PrOH, maximizing intermolecular interactions.

There are clear trends seen when comparing all of these alcohols, as compiled in Figure 8. The 1-D zigzag chains are hydrogen bonded networks in the √3 direction, closely following the underlying gold lattice spacing (similar growth has been seen with MeOH on Cu(111)[19]). The lateral spacing between adjacent chains in the close-packed surface direction is dependent on the strength of vdW forces. Increasing chain length, increases vdW contributions which compete with hydrogen bonding interactions. Larger alcohols tend to pack on the surface in progressively larger unit cells, unless geometric considerations increase hydrogen bonding (as with 2-PrOH). In Figure 8, the surface area per alcohol molecule can be compared to the alcohol's liquid phase volume. The 3-D volume ( $Å^3$ /molecule) is from the physical liquid density of each alcohol, it can be directly compared to the 2-D surface area ( $Å^2$ /molecule) as determined by the number of molecules per unit cell. The ML MeOH structure of zigzag pairwise chains (second column) is more complicated than the structures of the other alcohols, but can be simplified without the superstructure to the MeOH chain structure (third column), which can be compared with all other alcohols presented. Generally primary monohydroxy alcohols display a strong preference to form infinite hydrogen bonded chains in 3D crystalline form, while in secondary monohydroxy alcohols chains as well as rings can form[39,40]. In liquid studies it is generally found that primary alcohols form ever evolving chain-like aggregates, or cyclic structures[40], where zigzag chains are more preferred if steric factors allow. For MeOH, EtOH, PrOH and 2-PrOH chains dominate in solid crystals[41–43], while chains and cyclic rings are expected in liquid form[38,44–48]. But for 2-BuOH, helical chains form, a motif sometimes seen for secondary alcohols[39,49]. At the 2-D surface interface with Au(111); zigzag chains dominate the assembled structures of MeOH, EtOH, 1-PrOH and 2-PrOH; while, as in liquid, 2-BuOH forms more complex chiral long-range structures. Interestingly, we find that 3-D liquid density does not directly dictate 2-D surface packing density for these small alcohols. For EtOH and 2-PrOH, molecules take up the same surface area, while possessing different volumes in liquid. For the alternate case of 1-PrOH and 2-PrOH, very similar 3-D liquid volumes are not observed in 2D.

Taken together, the results of this study provide a framework for understanding the effect of the molecular structures of alcohols on the geometry of their surface-adsorbed structures. We predict linear alcohols with saturated alkyl tails $\geq$ C2 will adopt chain structures, as seen for ethanol. We also predict that, unlike in 3D liquids or solids, in alcohols $\geq$ C3, the position of the OH group along the alkyl chain will make a difference in 2D packing density due to the interplay of energies of the zigzag hydrogen bonded backbone with alkyl tail-tail van der Waals interactions. Furthermore, intrinsic chirality in the alkyl backbone is expected to influence the chirality of the surface adsorbed oxygen and preclude the possibility of zigzag chains that are composed of both surface bound chiralities of alcohol. These data and the interpretation offer fundamental insight and prediction of what may be expected for the self-assembly of other alcohols on metal surfaces.

**Acknowledgments/Funding**

This work was supported by the National Science Foundation under grant CHE-1412402 and CHE-1708397. ML thanks the NSF for a Graduate Research Fellowship. JC is supported by the MINECO through a Ramón y Cajal Fellowship and acknowledges support by the Marie Curie Career Integration Grant FP7-PEOPLE-2011-CIG: Project NanoWGS and The Royal Society through the Newton Alumnus scheme. A.M. is supported by the European Research Council under the European Union's Seventh Framework Programme (FP/2007-2013) / ERC Grant Agreement number 616121 (HeteroIce project). We are also grateful for computational resources to the London Centre for Nanotechnology, UCL Research Computing, and to the UKCP consortium (EP/ F036884/1) for access to Archer.